\documentclass[conference]{IEEEtran}
\IEEEoverridecommandlockouts

\usepackage{cite}
\usepackage{amsmath,amssymb,amsfonts}
\usepackage{algorithmic}
\usepackage{graphicx}
\usepackage{anyfontsize}
\usepackage{microtype}
\usepackage{mathtools}

\usepackage{textcomp}
\usepackage{xcolor}
\usepackage{placeins}
\usepackage{caption}
\usepackage{amsmath}
\usepackage{setspace}
\usepackage{relsize}

\usepackage{mathrsfs}
\usepackage[utf8]{inputenc}
\def\BibTeX{{\rm B\kern-.05em{\sc i\kern-.025em b}\kern-.08em
    T\kern-.1667em\lower.7ex\hbox{E}\kern-.125emX}}
\begin{document}
\title{{Joint Long-Term Processed Task and Communication Delay Optimization in UAV-Assisted MEC Systems Using DQN}\\
}
\author{\IEEEauthorblockN{Maryam Farajzadeh Dehkordi}
\IEEEauthorblockA{\textit{Department of Electrical and Computer Engineering} \\
\textit{George Mason University}\\
Fairfax, VA, USA \\
mfarajza@gmu.edu}
\and
\IEEEauthorblockN{ Bijan Jabbari}
\IEEEauthorblockA{\textit{Department of Electrical and Computer Engineering} \\
\textit{George Mason University}\\
Fairfax, VA, USA \\
bjabbari@gmu.edu}
}
\maketitle
\begin{abstract}
Mobile Edge Computing (MEC) assisted by Unmanned Aerial Vehicle (UAV) has been widely investigated as a promising system for future Internet-of-Things (IoT) networks. In this context, delay-sensitive tasks of IoT devices may either be processed locally or offloaded for further processing to a UAV or to the cloud. This paper, by attributing task queues to each IoT device, the UAV, and the cloud, proposes a real-time resource allocation framework in a UAV-aided MEC system. Specifically, aimed at characterizing a long-term trade-off between the time-averaged aggregate processed data (PD) and the time-averaged aggregate communication delay (CD), a resource allocation optimization problem is formulated. This problem optimizes communication and computation resources as well as the UAV motion trajectory, while guaranteeing queue stability. To address this long-term time-averaged problem, a Lyapunov optimization framework is initially leveraged to obtain an equivalent short-term optimization problem. Subsequently, we reformulate the short-term problem in a Markov Decision Process (MDP) form, where a Deep Q Network (DQN) model is trained to optimize its variables. Extensive simulations demonstrate that the proposed resource allocation scheme improves the system performance by up to 36\% compared to baseline models.
\end{abstract}


\section{Introduction}
The expansion of Internet of Things (IoT) technologies has sharply raised computational demands, particularly for low-latency applications such as video streaming, but many IoT devices remain limited in computing power \cite{Zarini2023URLLC}. Mobile Edge Computing (MEC) system enables these devices to offload their computationally-intensive tasks to nearby edge nodes with superior processing capabilities \cite{Tajallifar2022Energy}. Unmanned Aerial Vehicles (UAVs) further enhance MEC system by offering mobility, ease of deployment, and broad accessibility \cite{Jalali2024Energy}. However, due to the limited computing power of UAVs, cloud computing is also considered, though it introduces significant transmission delays because of the large distance between IoT devices and cloud servers. Driven to address this, we propose employing UAVs both as computational units and as relays via a backhaul link to the cloud, enabling efficient task offloading and optimizing the computational process.

In this context, extensive research has been conducted on UAV-assisted MEC systems \cite{Deep2022Zhang, Poursiami2024Multi, Optimizing2021Zhang, Latency2020Zhang, Online2021Hu, Stochastic2019Zhang, Dynamic2022Yang, Lyapunov2023Wu, Lyapunov2024Wang, Contract2021Lv, DRL2023Ma, Task2022Niu, Toward2020Wan}. While existing studies have provided valuable insights from an instantaneous perspective on various performance criteria such as quality of experience \cite{Deep2022Zhang}, energy consumption \cite{Poursiami2024Multi}, operational cost \cite{Optimizing2021Zhang}, and delay \cite{Latency2020Zhang}, they often overlook the long-term challenges posed by the non-deterministic conditions inherent in UAV-assisted communication networks. To address these challenges, recent research has explored UAV-enabled communication networks from a long-term causality perspective \cite{Online2021Hu, Stochastic2019Zhang, Dynamic2022Yang, Lyapunov2023Wu, Lyapunov2024Wang, Contract2021Lv}. After incorporating Lyapunov optimization techniques, approaches such as \cite{Online2021Hu, Stochastic2019Zhang, Dynamic2022Yang} have employed successive convex approximation, while others have designed heuristic algorithms \cite{Lyapunov2023Wu, Lyapunov2024Wang, Contract2021Lv}. However, traditional algorithms often struggle in environments that demand rapid decision-making, as they tend to be time-consuming, computationally intensive, and less adaptable. The dynamic and complex nature of UAV-assisted MEC systems requires optimization strategies that are not only efficient but also highly adaptive. In this regard, Deep Reinforcement Learning (DRL) emerges as a powerful alternative, capable of making effective sequential decisions under uncertainty \cite{Faramarzi2024Meta}.

This paper studies a three-tier UAV-aided MEC system, involving the cloud, UAV, and IoT devices that manage data queues for computing tasks. Our work emphasizes the long-term performance of UAV-assisted communication networks under uncertain conditions, unlike previous studies focused on instantaneous metrics \cite{Deep2022Zhang, Poursiami2024Multi, Optimizing2021Zhang, Latency2020Zhang}. Additionally, we introduce a novel metric, Processed Data Efficiency (PDE), which measures the ratio of Processed Data (PD) to Communication Delay (CD), offering a comprehensive evaluation of system performance. Furthermore, to ensure long-term system stability, we employ Lyapunov optimization, as done in prior works \cite{Online2021Hu, Stochastic2019Zhang, Dynamic2022Yang, Lyapunov2023Wu, Lyapunov2024Wang, Contract2021Lv}, but unlike their complex algorithms, we reformulate the problem as a Markov Decision Process (MDP) and use a Deep Q Network (DQN) model for real-time resource allocation. This reformulation enables more efficient and adaptive decision-making in dynamic environments. Extensive simulations demonstrate that the proposed resource allocation scheme improves the system performance by up to 36\% compared to baseline models.

\section{System Setup and Resource Allocation}\label{sec:System Setup and Resource Allocation}
\begin{figure}
    \centering
    \captionsetup{justification=centering}
    \includegraphics[width=1\linewidth]{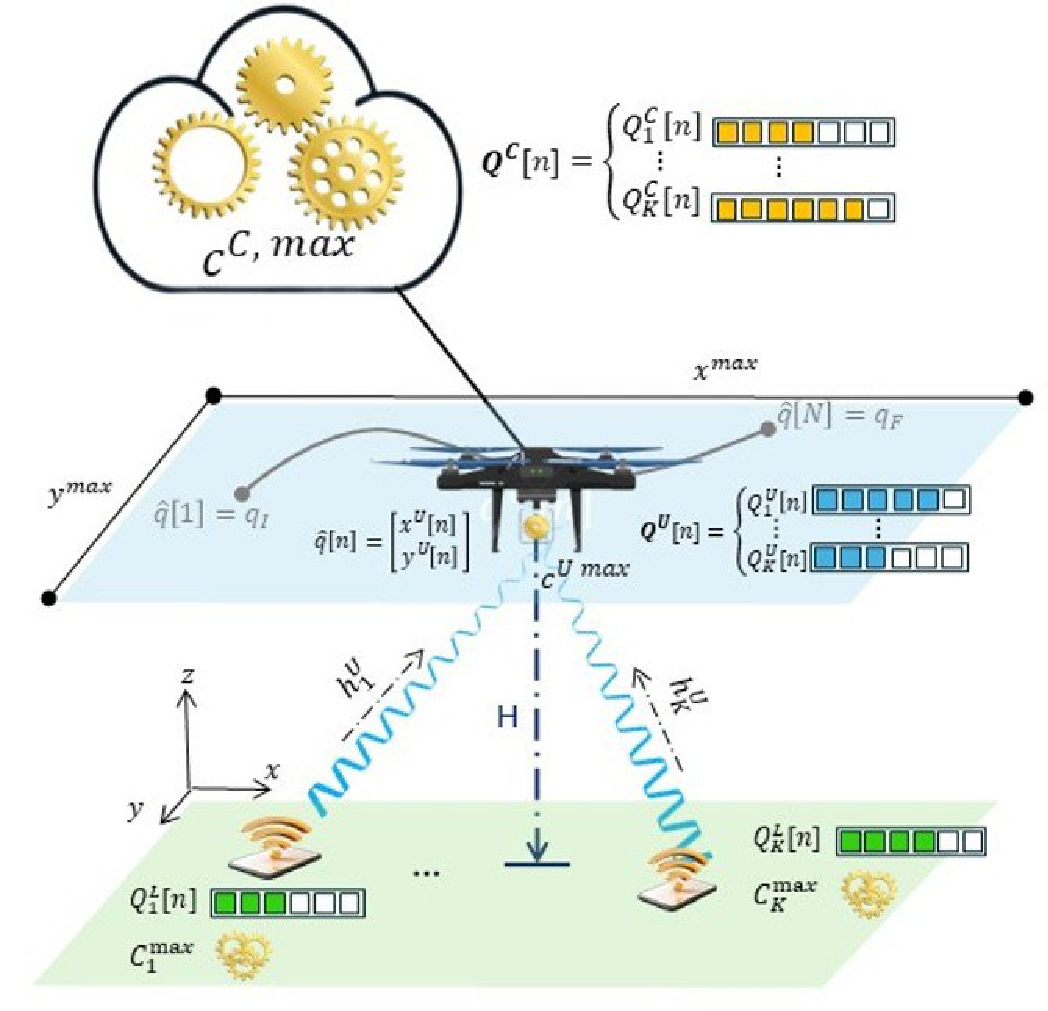}
    \caption{System model}
    \label{fig:system_model}
\end{figure}

As portrayed in Fig.~\ref{fig:system_model}, consider a MEC system with a set $\mathcal{K}=\{1, 2, \dots, K\}$ of $K$ IoT devices, equipped with limited computing resources that require an efficient offloading mechanism to outsource their computing tasks. In pursuit of this goal, the MEC network is equipped with a cloud, benefiting from large computing servers for processing the offloaded tasks of IoT devices. Additionally, a UAV facilitates the task offloading process, not only by acting as a relay to re-offload the computing tasks to the cloud, but also by processing part of them. More precisely, a three-tier computing regime is considered in the MEC system. First, a portion of each IoT device's task is executed locally at the IoT device. Next, a portion of the remaining task is offloaded to the UAV for processing. The UAV attempts to process part of the received task and finally re-offloads the remaining part to the cloud for processing. The communication link between the UAV and the cloud is assumed to be flat with sufficient bandwidth. All devices in the MEC network are assumed to be equipped with a single antenna. The UAV's mission is to fly from an appointed initial position to a final one. The flight duration is discretized into a set $\mathcal{N}=\{1, 2, \dots, N\}$ of ${N}$ intervals, each with equal duration $\tau$. To facilitate resource allocation, the interval duration is chosen to be sufficiently small, ensuring that the wireless channels remain quasi-static within each interval. We also assume that perfect channel state information is available in this system through existing channel estimation mechanisms~\cite{Secure2022Lu}.
\subsection{Channel Model}
Let $\mathbf{q}_k = \big[x_k, y_k\big]^T$ represent the position of the \(k^{\text{th}}\) IoT device. In the three-dimensional Cartesian coordinate system, the horizontal position of the UAV at the \(n^{\text{th}}\) interval is denoted by $\mathbf{\hat{q}}[n] = \big[x^{\text{U}}[n], y^{\text{U}}[n]\big]^T$, where the UAV is assumed to fly at a constant altitude \(H\). By assuming the height of all IoT devices to be zero, the Euclidean distance between the UAV and the \(k^{\text{th}}\) IoT device at the \(n^{\text{th}}\) interval is calculated as:
\begin{align} \label{eq:distance}
d_k^{\text{U}}[n] = \sqrt{||\mathbf{\hat{q}}[n] - \mathbf{q}_k||^2 + H^2}.
\end{align}
For ease of resource allocation, this paper considers a block fading channel model, where the channel condition is stable within an interval but may vary independently between consecutive intervals. The uplink channels are assumed to experience large-scale path loss fading and independent, identically distributed small-scale Rician fading. For the \(k^{\text{th}}\) IoT device at any given \(n^{\text{th}}\) interval, the channel gain is modeled as:
\begin{align} \label{eq:channel_gain}
h_k^{\text{U}}[n] = \sqrt{\eta_k^{\text{U}}[n]}\rho_k^{\text{U}}[n],
\end{align}
as in \cite{3D2019You}, where $\eta_k^{\text{U}}[n]$ denotes the large-scale average channel power gain, accounting for the path loss, which is given by:
\begin{align} \label{eq:channel_coeff}
\eta_k^{\text{U}}[n] = \eta_0\left(d_k^{\text{U}}[n]\right)^{-\theta} = \frac{\eta_0}{\left(d_k^{\text{U}}[n]\right)^{\theta}}.
\end{align}Here, $\eta_0$ refers to the average channel power gain at the reference distance \(d_0 = 1\) m, and $\theta$ is the path loss exponent. Additionally, $\rho_k^{\text{U}}[n]$ accounts for small-scale fading. Due to the maneuverability of the UAV, it is reasonable to assume that Line of Sight (LoS) links are established. Therefore, we model the small-scale fading between the UAV and the \(k^{\text{th}}\) IoT device using Rician fading as follows:
 \begin{align} \label{eq:LineOfSight}
\rho_k^{\text{U}}[n]=\underbrace{\sqrt{\frac{K^{\text{R}}}{K^{\text{R}}+1}}\rho_k^{\text{LoS}}[n]}_{\textrm{LoS Link }}+\underbrace{\sqrt{\frac{1}{K^{\text{R}}+1}}\rho_k^{\text{NLoS}}[n]}_{\textrm{NLoS Links }},
\end{align}in which the LoS link includes a dominant path directing between the transceivers, with $\rho_k^{\text{LoS}}$ expressing the deterministic LoS channel component, such that $|\rho_k^{\text{LoS}}|=1$. The Non-Line of Sight (NLoS) component $\rho_k^{\text{NLoS}}$ follows a distribution with zero mean and unit variance. Finally, $K_k^{\text{U}}[n]$ represents the Rician factor of the channel between $k^{\text{th}}$ IoT device and the UAV at $n^{\text{th}}$ interval.
\subsection{Communication Model}
Once all IoT devices offloaded their tasks in uplink, the received signal by the UAV at $n^{\text{th}}$ interval is formulated as:
\begin{equation}
   y^{\text{U}}[n]=\sum\nolimits_{k \in \mathcal{K}}h_k^{\text{U}}[n]\sqrt{p_{k[n]}}u_{k}[n]+z^{\text{U}},
\end{equation}
in which \(u_k[n]\) and \(p_k[n]\) represent the unit-power task information symbol and the uplink transmit power of the \(k^{\text{th}}\) IoT device at the \(n^{\text{th}}\) interval, respectively. Additionally, \(z^{\text{U}} \sim \mathcal{CN}(0, \sigma_{z}^{2})\) specifies the zero-mean Additive White Gaussian Noise (AWGN) with variance \(\sigma_{z}^{2}\) at the UAV. Consequently, the Signal-to-Interference-plus-Noise-Ratio (SINR) from the transmission of the \(k^{\text{th}}\) IoT device, as received by the UAV at the \(n^{\text{th}}\) interval, is expressed as:
 \begin{align}
\gamma_k[n]=\frac{\left\vert {h}_k^{\text{U}}[n]\sqrt{p_{k}[n]}\right\vert ^{2}}{\sum\nolimits\limits_{i = 1,i \ne k}^K \left\lvert {h}_k^{\text{U}}[n]\sqrt{p_i[n]} \right\rvert^{2}+\sigma_{z}^{2}}.
\end{align}Accordingly, the maximum achievable data rate received by the UAV from the transmission of $k^{\text{th}}$ IoT device at $n^{\text{th}}$ interval according to the Shannon-Hartley theorem is calculated as:
 \begin{align}
R_{k}[n]=B_0\log _{2}\left( 1+\gamma _{k}[n]\right),
\end{align}where $B_0$ denotes the communication bandwidth in Hertz. Let $D^{\text{U}}_{k}[n]$ denote the data bits offloaded to the UAV by $k^{\text{th}}$ IoT device at $n^{\text{th}}$ interval at a data rate of $R_{k}[n]$. Accordingly, the CD for transmission of $k^{\text{th}}$ IoT device to the UAV at $n^{\text{th}}$ interval is calculated as:
 \begin{align} \label{eq:time_comm}
t_{k}^{\text{U, Comm}}[n] = \frac{D^{\text{U}}_{k}[n]}{R_k[n]}.
\end{align}Further, let $t_{k}^{\text{C, Comm}}[n]$ denote the communication  delay corresponding to $k^{\text{th}}$ IoT device's task offloaded from the UAV to the cloud at $n^{\text{th}}$ interval is calculated as:
 \begin{align} \label{eq:time_comp_c}
 t_{k}^{\text{C, Comm}}[n] = 
    \begin{cases}
        t_0, & x_k^{\text{C}}[n] > 0, \\
        0, & \text{otherwise},
    \end{cases}
\end{align}with $x_k^{\text{C}}[n]$, representing the cloud's offloading indicator for the $k^{\text{th}}$ IoT device's task, indicating the portion of that task that is to be re-offloaded from the UAV to the cloud at $n^{\text{th}}$ interval and $t_0$, representing the installation delay required to establish high-data-rate communication between the UAV and the cloud. Consequently, the overall CD for $k^{\text{th}}$ IoT device's task at $n^{\text{th}}$ interval denoted by $t_k^{\text{Comm}}[n]$ includes the CD from the IoT device to the UAV and that from the UAV to the cloud, calculated as:
 \begin{align}\label{eq:comm_delay}
 t_k^{\text{Comm}}[n]=t_{k}^{\text{U, Comm}}[n]+t_{k}^{\text{C, Comm}}[n].
\end{align}
\subsection{Computation Model}
Due to limited computing capacity on IoT devices and the UAV, tasks may be offloaded first to the UAV and then to the cloud for complete processing. In line with this, let us denote the Central Processing Unit (CPU) frequency allocated to the task of $k^{\text{th}}$ IoT device at $n^{\text{th}}$ interval locally by that IoT device, the UAV, and the cloud, in cycle per second, by $c^{\text{L}}_k[n]$, $c^{\text{U}}_k[n]$, and $c^{\text{C}}_k[n]$, respectively. Accordingly, the computation delay at the local device, the UAV, and the cloud for $k^{\text{th}}$ IoT device at $n^{\text{th}}$ interval are denoted by $t_{k}^{\text{L, Comp}}[n]$, $t_{k}^{\text{U, Comp}}[n]$, and $t_{k}^{\text{C, Comp}}[n]$, respectively, among which $t_{k}^{\text{L, Comp}}[n]$ is calculated as follows:
 \begin{align} \label{eq:time_comp}
t_{k}^{\text{L, Comp}}[n] = \frac{B_{k}^{\text{L}}[n]}{c_k^{\text{L}}[n]}C,
\end{align}
where $B_{k}^{\text{L}}[n]$ refers to the total number of bits of $k^{\text{th}}$ IoT device's task at $n^{\text{th}}$ interval allocated to be processed locally, and $C$ denotes the number of CPU frequency cycles required to process one bit of each IoT device's task, measured in CPU cycles per bit. Similarly, we can calculate $t_{k}^{\text{U, Comp}}[n]$ and $t_{k}^{\text{C, Comp}}[n]$. It is presumed that each IoT device performs local computing and task offloading at the same time, as does the UAV. As such, the total delay for processing $k^{\text{th}}$ IoT device's task at $n^{\text{th}}$ interval measured in seconds is calculated as:
 \begin{align}
    T_k^{\text{Tot}}[n] &= \text{max}\{t_{k}^{\text{L, Comp}}[n], t_{k}^{\text{U, Comm}}[n]\} \nonumber \\
    &\quad + \text{max}\{t_{k}^{\text{U, Comp}}[n], t_{k}^{\text{C, Comm}}[n]\} + t_{k}^{\text{C, Comp}}[n].
\end{align}
\subsection{Queue Dynamics and Stability}
Each IoT device incorporates a task arrival queue to buffer arrival tasks for processing. Moreover, the UAV and the cloud are each equipped with $K$ queues, each of which is associated with the task of one IoT device. We assure network stability based on the queue backlog at all IoT devices, the UAV and the cloud. Let $\mathbf{I}[n] = [I_k[n]]$ represent the arrival of computing task bits for IoT devices at $n^{\text{th}}$ interval, where $I_{k}[n]$ specifies the random task arrival of $k^{\text{th}}$ IoT devices at $n^{\text{th}}$ interval. This rate is assumed to follow independent and identically distributed (i.i.d.) with the budget of $I_{k}^{\text{max}}$ for  $k^{\text{th}}$ IoT device where $k \in \mathcal{K}$. Let $Q_{k}^{\text{L}}[n]$ introduce the task queue backlog, maintained at $k^{\text{th}}$ IoT device at $n^{\text{th}}$ interval, with $Q_{k}^{\text{L}}[0]=0$. Then, we express the portion of the queue backlog, offloaded to the UAV at $n^{\text{th}}$ interval as $D_{k}^\text{U}[n]$, which is calculated by:
\begin{align}\label{eq:off_bits}
D_{k}^{\text{U}}[n]=x_{k}^{\text{U}}[n]Q_{k}^{\text{L}}[n], 
\end{align}
whereas from the remaining queue, the portion that is already being processed locally by $k^{\text{th}}$ IoT device at $n^{\text{th}}$ interval, denoted by $B_k^{\text{L}}[n]$ is calculated as:
\begin{align}\label{eq:off_bits1}
B_k^{\text{L}}[n]=w_{k}^{\text{L}}[n](1-x_{k}^{\text{U}}[n])Q_{k}^{\text{L}}[n].
\end{align}
Here, $w_{k}^{\text{L}}[n]$ corresponds to the portion decision of remaining data bits which is to be processed at $k^{\text{th}}$ IoT device's queue at $n^{\text{th}}$ interval. Given this, the task queue of $k^{\text{th}}$ IoT device can be updated as:
\begin{align}\label{eq:Q_queue_L}
Q_{k}^{\text{L}}[n+1]=\{Q_{k}^{\text{L}}[n]-D_{k}^{\text{U}}[n]-B_{k}^{\text{L}}[n]\}^{+}+I_{k}[n],
\end{align}in which $\{f\}^{+}$ indicates $\max \{0,f\}$ to guarantee queue causality. At the UAV, the task queue backlog $Q_{k}^{\text{U}}[n]$, which is dedicated for buffering the tasks of $k^{\text{th}}$ IoT device at $n^{\text{th}}$ interval, satisfies the following update process:
 \begin{align}\label{eq:Q_queue_U}
Q_{k}^{\text{U}}[n+1]=\{Q_{k}^{\text{U}}[n]-D_{k}^{\text{C}}[n]-B_{k}^{\text{U}}[n]\}^{+}+D_{k}^{\text{U}}[n],
\end{align} where $D_{k}^{\text{C}}[n]$ denotes the data bits offloaded to the cloud and calculated similar to \eqref{eq:off_bits}. Note that for the remaining data corresponding to the \(k^{\text{th}}\) IoT device's task at the \(n^{\text{th}}\) interval in the UAV's queue, i.e., \((1 - x_k^{\text{C}}[n])Q_k^{\text{U}}[n]\), it is assumed that a portion, $w_k^{\text{U}}[n]$, is already being processed by the UAV, denoted by $B_k^{\text{U}}[n]$, and calculated similarly to \eqref{eq:off_bits1}. Similar to the UAV, the task queue backlog $Q_{k}^{\text{C}}[n]$ at the cloud for buffering $k^{\text{th}}$ IoT device's task at $n^{\text{th}}$ interval updates as beneath:
 \begin{align}\label{eq:Q_queue_c}
Q_{k}^{\text{C}}[n+1]=\{Q_{k}^{\text{C}}[n]-B_{k}^{\text{C}}[n]\}^{+}+D_{k}^{\text{C}}[n],
\end{align}where $B_{k}^{\text{C}}[n]$ corresponds to the data bits of $k^{\text{th}}$ IoT device at $n^{\text{th}}$ interval processed at the cloud, given by:
 \begin{align}\label{eq:B_queue_c}
B_{k}^{\text{C}}[n]=w_k^{\text{C}}[n]Q_{k}^{\text{C}}[n].
\end{align}
Here, $w_{k}^{\text{C}}[n]$ signifies the portion of data bits of $k^{\text{th}}$ IoT device's queue at $n^{\text{th}}$ interval, which is already being processed by the cloud. Consequently, each IoT device needs to guarantee that its queue is stable locally according to \cite{neely2022stochastic}:
 \begin{align}\label{eq:Q_stability}
\bar{Q}_{k}^\text{L} = \lim_{N \rightarrow \infty }\frac{1}{N}\sum\nolimits_{n \in \mathcal{N}}{\mathbb{E}\left\{Q_{k}^\text{L}[n]\right\} } < \infty, \quad k \in \mathcal{K},
\end{align}where $Q_{k}^{\text{L}}[n]\geq 0$. Stability for both UAV and cloud queues is ensured in a similar manner.
\subsection{Problem Formulation}\label{subsec: Problem Formulation}
We consider a trade-off between the amount of PD and the CD in the UAV-assisted MEC network. Many existing works optimize the instantaneous performance of MEC systems \cite{Deep2022Zhang, Latency2020Zhang, Optimizing2021Zhang} while a long-term optimization would offer a better understanding of their performance. In light of this, we focus on characterizing the trade-off between the long-term PD and the long-term CD in the UAV-assisted MEC network, referred to as the PDE. Toward this goal, let $B_k^{\text{Tot}}[n]$ define the total amount of the PD for $k^{\text{th}}$ IoT device at $n^{\text{th}}$ interval, which is calculated as:
 \begin{align}\label{eq:b_tot}
 B_k^{\text{Tot}}[n] = \sum\nolimits_{k\in \mathcal{K}} B_{k}^{\text{L}}[n]+B_{k}^{\text{U}}[n]+B_{k}^{\text{C}}[n].
\end{align}
Now, the long-term PDE is formally defined as:
 \begin{align}\label{eq:utility_func}
U = \frac{\underset{N\to \infty}{\lim}{\frac{1}{N} \displaystyle\sum\nolimits_{n \in \mathcal{N}} \displaystyle\sum\nolimits_{k \in \mathcal{K}}B_k^{\text{Tot}}[n]}}{\underset{N \to \infty}{\lim} {\frac{1}{N} \displaystyle\sum\nolimits_{n \in \mathcal{N}}\displaystyle\sum\nolimits_{k \in \mathcal{K}}t_k^{\text{Comm}}[n]}}.
\end{align}
For each IoT device's task at each interval, i.e., $\forall k \in \mathcal{K}, \forall n \in \mathcal{N}$, we optimize the system parameters, including the uplink transmit power of IoT devices, $\mathbf{P} = \left[p_k[n]\right]$, the offloading indicator $\mathbf{X} = \left[\mathbf{x}_k[n]\right]$ where $\mathbf{x}_k[n] = \big[x_k^\text{U}[n], x_k^\text{C}[n]\big]$, the computing indicator, $\mathbf{W} = \left[\mathbf{w}_k[n]\right]$ where $\mathbf{w}_k[n] = \big[w_k^\text{L}[n], w_k^\text{U}[n], w_k^\text{C}[n]\big]$, the computing resource allocation, $\mathbf{C} = \left[\mathbf{c}_k[n]\right]$ where $\mathbf{c}_k[n] = \big[c_k^\text{L}[n], c_k^\text{U}[n], c_k^\text{C}[n]\big]$, and the UAV motion trajectory, $\mathbf{\hat{q}}[n] = \big[x^{\text{U}}[n], y^{\text{U}}[n]\big]$. The problem is then formulated as follows:
\begin{align}
\mathscr{P}_{1}: &\underset{\textbf{P, X, W, C, $\mathbf{\hat{q}}$}}{\max}\quad {U} \nonumber\\
s.t. \quad & \text{C}_1: 0 \leq p_{k}[n] \leq {P}^{\text{max}}, \quad \forall n \in \mathcal{N}, \nonumber\\
&\text{C}_2: \ 0 \leq x_{k}^{i}[n] \leq 1, \quad  \forall i \in \{\text{U, C}\},k \in \mathcal{K}, \forall n \in \mathcal{N},\nonumber\\
&\text{C}_3: \ 0 \leq w_{k}^{i}[n] \leq 1, \quad  \forall i \in \{\text{L, U, C}\}, \forall k \in \mathcal{K}, \forall n \in \mathcal{N}, \nonumber\\
&\text{C}_4: \ 0 \leq c_{k}^{\text{L}}[n] \leq {C}_{k}^{\text{max}}, \quad  \forall k \in \mathcal{K}, \forall n \in \mathcal{N}, \nonumber\\
&\text{C}_5: \ 0 \leq \sum\nolimits_{k \in \mathcal{K}}c_{k}^{\text{U}}[n] \leq {C}^{\text{U, max}}, \quad \forall n \in \mathcal{N},\nonumber\\
&\text{C}_6: \ 0 \leq \sum\nolimits_{k \in \mathcal{K}}c_{k}^{\text{C}}[n] \leq {C}^{\text{C, max}},  \quad \forall n \in \mathcal{N},\nonumber\\
&\text{C}_7: 0 \leq \frac{\ \left\|\mathbf{\hat{q}}[n+1] - \mathbf{\hat{q}}[n]\right\|}{\tau} \leq V^{\text{max}} ,\quad  \forall n \in \mathcal{N},    \nonumber\\
&\text{C}_8: \mathbf{\hat{q}}[1] = \mathbf{q}_I, \ \mathbf{\hat{q}}[N] = \mathbf{q}_F,\nonumber\\
&\text{C}_9: \mathbf{\hat{q}}[n] \in \mathcal{A}^\text{UAV}, \forall n \in \mathcal{N},\nonumber\\
&\text{C}_{10}: \ T_k^{\text{Tot}}[n] \leq \tau, \quad \forall k \in \mathcal{K}, \nonumber\\
&\text{C}_{11}: \bar{Q}_k^{i} < \infty, \quad i \in \{\text{L, U, C}\}, \forall k \in \mathcal{K}.
\label{eq:em10}
\end{align} Here, $\text{C}_1$ ensures that the IoT devices operate within their transmit power budget at all intervals; $\text{C}_2$ and $\text{C}_3$, specify the portion of offloading and computing indicators; $\text{C}_4$ assures the computing capacity of $k^{\text{th}}$ IoT device, up to $\text{C}_{k}^{\text{max}}$; The computing resources allocated to the tasks at UAV and cloud are bounded to $\text{C}^{\text{U, max}}$ and $\text{C}^{\text{C, max}}$ in $\text{C}_5$ and $\text{C}_6$, respectively; $\text{C}_7$ restricts the flying velocity of the UAV, up to $V^{\text{max}}$; $\text{C}_8$ ensures that the UAV eventually returns to its initial flying location, with $\mathbf{q}_I$ and $\mathbf{q}_F$, being the initial and final location of the UAV; $\text{C}_9$ determines the horizontal area denoted by $\mathcal{A}^\text{UAV}$ that the UAV can fly. $\text{C}_{10}$ limits the total delay for tasks of IoT devices up to $\tau$ and finally, $\text{C}_{11}$ guarantees for the computing queue stability in IoT devices, the UAV, and the cloud, respectively.

Observably, the coupling of continuous decision variables $\mathbf{P, X, W, C}$ and the integer decision variable $\hat{q}$ makes problem $\mathscr{P}_{1}$ a mixed-integer non-linear programming (MINLP) problem, which is NP-hard. While a brute-force exhaustive search could find a globally optimal solution, it becomes impractical due to the problem's complexity and scalability, even for moderately sized systems. Previous approaches, based on convex optimization theory \cite{Online2021Hu, Stochastic2019Zhang, Dynamic2022Yang}, involved complex and time-consuming transformations to obtain locally optimal solutions. However, given the dynamic nature of the MEC network, driven by UAV mobility, a real-time solution is necessary. To address this, we reformulate the objective function using non-linear fractional programming, making it more tractable. As a result, the short-term PDE is computed as:
\begin{equation} \label{eq:pertimeslot}
U[n]=\frac{\displaystyle\sum\nolimits_{k=1}^{K}\displaystyle\sum\nolimits_{m=0}^{n-1}{B}_k^{\text{Tot}}[m]}{\displaystyle\sum\nolimits_{k=1}^{K}\displaystyle\sum\nolimits_{m=0}^{n-1} {t}_k^{\text{Comm}}[m]},\quad \forall n \in \mathcal{N}.
\end{equation} 
Next, we apply the drift-plus-penalty to solve stochastic optimization problems using Lyapunov optimization. This transforms the original problem $\mathscr{P}{1}$ into a per-interval optimization problem $\mathscr{P}{2}$, where a constant $V$ balances queue stability and system utility, based only on current queue lengths as follows:
 \begin{align} \label{eq:problem_3}
\mathscr{P}_{2}:&\max_{\mathbf{P}, \mathbf{X}, \mathbf{W}, \mathbf{C}, \mathbf{\hat{q}}} V\left(\displaystyle\sum\nolimits_{k \in \mathcal{K}}{B}_k^{\text{Tot}}[n]-U[n]\displaystyle\sum\nolimits_{k \in \mathcal{K}} {t}_k^{\text{Comm}}[n]\right)\nonumber  \\
&+\sum\nolimits_{k \in \mathcal{K}} \left (Q_{k}^{\text{L}}[n](B_{k}^{\text{L}}[n]+D_{k}^{\text{U}}[n])\right)  \nonumber \\
&-\sum\nolimits_{k \in \mathcal{K}}\left(Q_{k}^{\text{U}}[n](D_{k}^{\text{U}}[n]-B_{k}^{\text{U}}[n]-D_{k}^{\text{C}}[n])\right) \nonumber \\
&-\sum\nolimits_{k \in \mathcal{K}}\left (Q_{k}^{\text{C}}[n]( D_{k}^{\text{C}}[n]-B_{k}^{\text{C}}[n])\right) \nonumber \\
\text{s.t.} & \quad \text{C}_1 - \text{C}_{10} \quad in \quad \mathscr{P}_{1},
\end{align}
The full proof has been provided in greater detail in \cite{dehkordi2024deep}.

As seen, \eqref{eq:problem_3} is now a per-interval optimization problem that can be modeled as a MDP. An MDP consists of a state space, an action space, and a reward function. The state space $\mathcal{S}$ includes all possible states at each interval. At the \(n^{\text{th}}\) interval, the state is denoted by \(\mathbf{s}[n] \in \mathcal{S}\) and defined as $\mathbf{s}[n] = \{s_k[n], \forall k \in \mathcal{K}\}$ where \(s_k[n]\) represents the state of the \(k^{\text{th}}\) IoT device, given by $s_k[n] = \big[Q_k^\text{L}[n], Q_k^\text{U}[n], Q_k^\text{C}[n]\big]$. The action space $\mathcal{A}$ consists of all possible actions for each state. At the \(n^{\text{th}}\) interval, the action is denoted by \(\mathbf{a}[n] \in \mathcal{A}\) and defined as $\mathbf{a}[n] = \{a_k[n], \forall k \in \mathcal{K}\}$ where \(a_k[n]\) signifies the action of the \(k^{\text{th}}\) IoT device, given by $a_k[n] = \big[p_k[n], \mathbf{x}_k[n], \mathbf{w}_k[n], \mathbf{c}_k[n], \hat{\mathbf{q}}[n]\big]$. The reward function, \(r(\mathbf{s}[n], \mathbf{a}[n]) : \mathcal{S} \times \mathcal{A} \rightarrow \mathbb{R}\), maps states and actions to a real-valued reward. The overall reward at the \(n^{\text{th}}\) interval is given by $r(\mathbf{s}[n], \mathbf{a}[n]) = \sum_{k \in \mathcal{K}} r_k(s_k[n], a_k[n])$,
where \(r_k(s_k[n], a_k[n])\) is the reward for the \(k^{\text{th}}\) IoT device based on its state \(s_k[n]\) and action \(a_k[n]\), in which:
 \begin{align} \label{eq:reward}
& r_k\big({s}_k[n],{a}_k[n]\big) = v_1 \bigg( Q_{k}^{\text{L}}[n]\left(B_{k}^{\text{L}}[n]+D_{k}^{\text{U}}[n]\right) \bigg) \nonumber \\
&-v_2 \bigg( Q_{k}^{\text{U}}[n]\left(D_{k}^{\text{U}}[n]-B_{k}^{\text{U}}[n]-D_{k}^{\text{C}}[n]\right) \bigg) \nonumber \\
&-v_3 \bigg(Q_{k}^{\text{C}}[n]\left( D_{k}^{\text{C}}[n]-B_{k}^{\text{C}}[n]\right) \bigg) \\
&+v_4 \bigg( \big( \displaystyle{B}_k^{\text{Tot}}[n] - \displaystyle{t}_k^{\text{Comm}}[n] \frac{\sum\nolimits_{m=0}^{n-1}{B}_k^{\text{Tot}}[m]}{\sum\nolimits_{m=0}^{n-1}{t}_k^{\text{Comm}}[m]}  \big) \eta_k[n] \bigg) \nonumber.
\end{align}Here, $v_1$, $v_2$, $v_3$, and $v_4$ are experimental parameters used to balance the contribution of each factor in the reward function, while $\eta_k[n]$ indicates whether condition $\text{C}_{10}$ in problem $\mathscr{P}{1}$ is violated (with a value of 0) or satisfied (with a value of 1). At the \(n^{\text{th}}\) interval $n \in \mathcal{N}$, the agent observes the current state of the environment and takes an action based on its policy. A reward, determined by the current state and the agent’s action, is then provided, and the environment transitions to the next state, \(\textbf{s}[n+1]\). The objective is to learn a policy \(\pi(\textbf{a}[n]|\textbf{s}[n]):\mathcal{S} \rightarrow \mathcal{A}\) that maximizes the accumulated expected reward.
\section{DQN Approach}\label{sec:Deep Q-Network}
With the agent serving as the resource allocator for the network, a DRL is designed to optimize the trade-off between queue dynamics and PDE. The effectiveness of this algorithm relies on a well-structured reward function and carefully defined state and action spaces, as previously outlined. To validate the feasibility of the approach, we selected the DQN for its simplicity. DQN integrates state, action, and reward to maximize a cumulative discounted reward, which is approximated by a neural network \cite{powell2007approximate}. Key techniques, including experience replay and fixed target networks, ensure stability during training, allowing the agent to learn near-optimal policies in complex environments. Further details on the DQN framework are available in \cite{dehkordi2024deep}.
\section{Numerical Results}

In this section, we evaluate the feasibility of the proposed resource allocation scheme through comprehensive numerical simulations conducted using Python 3.7. The UAV's trajectory begins and ends at the center of a coverage area measuring 500×500 $\text{m}^2$, where two IoT devices are randomly positioned. Each simulation interval, denoted as \(\tau\), spans 1 second, with data arrival for each IoT device per interval randomly distributed between 0 and \(2.5 \times 10^5\) bits. The remaining simulation parameters are as follows: \({P}^{\text{max}} = 0.1\, \text{W}\), \({C}^{\text{U, max}} = 5 \times 10^6\, \text{cycles/sec}\), \({C}^{\text{C,max}} = 10^8\, \text{cycles/sec}\), \(N = 1000\), \(H = 100\, \text{m}\), \(\eta_0 = -40\, \text{dB}\), \({V}^{\text{max}} = 30\, \text{m/sec}\), \(\theta = 2\), \(K^{\text{R}} = 10\), \(\sigma_z^2 = -174\, \text{dBm}\), \({B}_0 = 180\, \text{KHz}\), \(C = 1024\, \text{cycles/bit}\), and \(t_0 = 0.25\, \text{sec}\). For simplicity, the action space is restricted to include the UAV and cloud offloading indicators (denoted as \(x_k^{\text{U}}[n]\) and \(x_k^{\text{C}}[n]\), respectively), processing levels at both the UAV and the cloud (denoted as \(w_k^{\text{U}}[n]\) and \(w_k^{\text{C}}[n]\), respectively), and the UAV’s trajectory coordinates (\(x^{\text{U}}[n]\), \(y^{\text{U}}[n]\)). Other optimization variables are predefined in the simulations. The actions are discretized into two levels: 30\% or 60\% of the data is offloaded to or processed by the UAV or the cloud. The state space of each data queue is divided into 10 levels. To evaluate the performance of the proposed setup, 1000 channel realizations were used.

Using these simulation parameters, we performed a series of evaluations to compare the performance of the proposed DQN algorithm with three baseline approaches. The proposed algorithm leverages DQN to allocate resources, optimizing both UAV trajectory and computation policies, while fixing the transmission power and computing resource allocation for simplicity. The Random baseline randomly offloads tasks between the UAV and the cloud, with a randomly generated UAV trajectory. In the UAV baseline, 60\% of tasks are allocated to the centrally located UAV. Similarly, in the Cloud baseline, 60\% of tasks are processed by the cloud, with the UAV centrally positioned.

\begin{figure}
    \centering
    \includegraphics[width=1\linewidth]{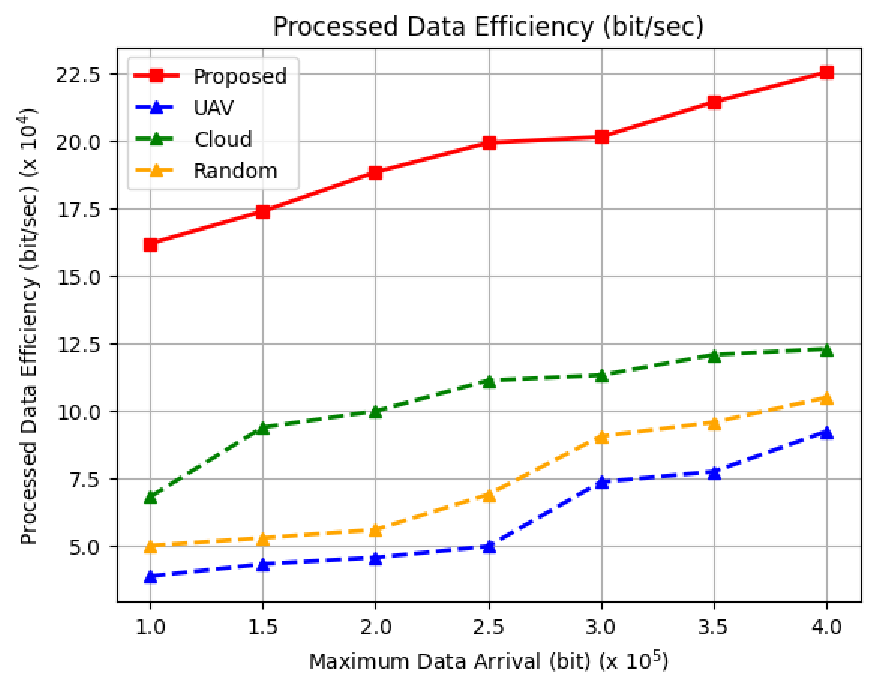}
    \caption{ PDE vs. maximum data arrival}
    \label{fig:be_arr}
\end{figure}

\begin{figure}
    \centering
    \includegraphics[width=1\linewidth]{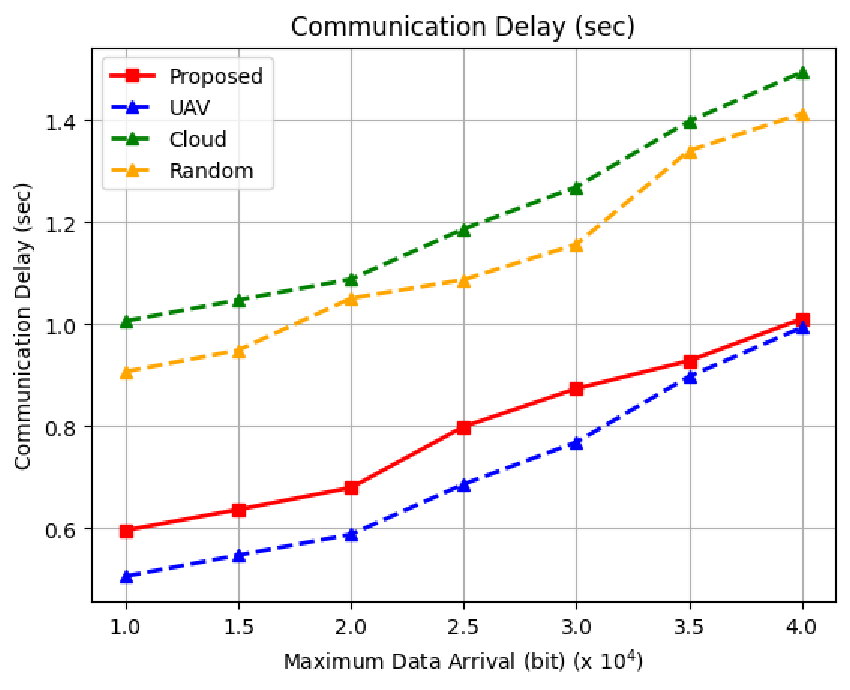}
    \caption{CD vs. maximum data arrival}
    \label{fig:d_arr}
\end{figure}

\begin{figure}
    \centering
    \includegraphics[width=1\linewidth]{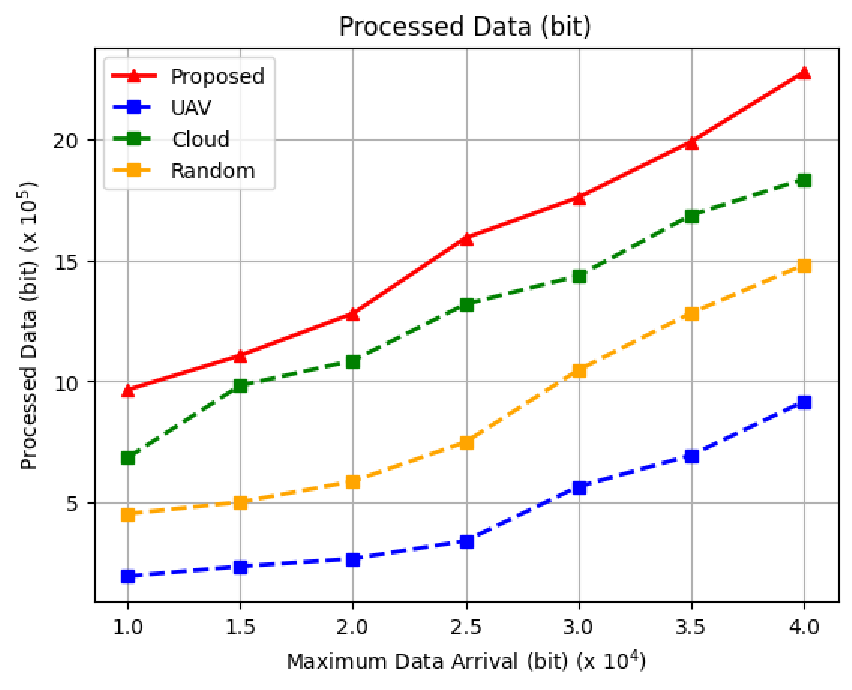}
    \caption{PD vs. maximum data arrival}
    \label{fig:b_arr}
\end{figure}

Figs. \ref{fig:be_arr} to \ref{fig:b_arr}, illustrate the impact of maximum arrival data (in bits, scaled by a afctor of $10^5$) on system PDE \eqref{eq:utility_func}, system CD \eqref{eq:comm_delay}, and system PD. Fig.~\ref{fig:be_arr} shows that as the maximum data arrival rate increases, system PDE improves across all baselines. The proposed strategy, optimizing trajectory and computing allocation, achieves the highest PDE. The Cloud baseline, relying solely on cloud processing, ranks second. The Random baseline outperforms the UAV-only baseline, which records the lowest PDE due to limited computing resources. Fig. \ref{fig:d_arr} demonstrates that increasing data arrival rates leads to higher system CD for all baselines. The UAV baseline exhibits the lowest CD due to its proximity to IoT devices, enabling faster data transfer. The Proposed baseline, which strategically allocates tasks between the UAV and the cloud, ranks second. The Random baseline, with its random task assignments, outperforms the Cloud baseline, which incurs the highest CD due to its remote processing location from the IoT devices.
\section{Conclusions}\label{sec:Conclusion and Future Work}
This paper investigates the long-term balance between PD and CD in a UAV-assisted MEC network, ensuring queue stability across IoT devices, the UAV, and the cloud. The proposed algorithm optimizes communication, computation, and UAV trajectory decisions, achieving a significant improvement in PDE, with up to a $36\%$. Although the CD doesn't match the lower delay of the UAV-only baseline, the superior PDE highlights the benefits of joint optimization. The preliminary work presented here demonstrates the feasibility of applying DQN to optimize the trade-off between PDE and network stability, and future work will be directed toward improving the algorithm's efficiency and effectiveness.

\bibliographystyle{IEEEtran}
\bibliography{AbbBib}
\end{document}